\documentclass[showpacs,prl,twocolumn,nofootinbib,preprintnumbers]{revtex4-1} 
\pdfoutput=1

\usepackage{graphicx}  
\usepackage{amssymb,amsmath}   
\usepackage{shuffle}
 \usepackage{url}
 \usepackage{hyperref}
\hypersetup{colorlinks=true,linkcolor=magenta,anchorcolor=green,citecolor=cyan,filecolor=black,menucolor=black,urlcolor=black}
 \usepackage[table]{xcolor}

\usepackage{feynmp-auto}

\newcommand{\pentg}[5]{
  \parbox{30pt}{
\begin{fmfgraph*}(30,30)
\fmfsurroundn{i}{5}
\fmf{plain}{i1,v1}
\fmf{plain}{i2,v2}
\fmf{plain}{i3,v3}
\fmf{plain}{i4,v4}
\fmf{plain}{i5,v5}
\fmf{plain,tension=0.5}{v1,v2,v3,v4,v5,v1}
\fmfv{label=${\color{blue}#4}$,label.dist=3pt}{i1}
\fmfv{label=${\color{blue}#3}$,label.dist=3pt}{i2}
\fmfv{label=${\color{blue}#2}$,label.dist=3pt}{i3}
\fmfv{label=${\color{blue}#1}$,label.dist=3pt}{i4}
\fmfv{label=${\color{blue}#5}$,label.dist=3pt}{i5}
\fmf{fermion,tension=0,label=$\ell$,label.side=left}{v5,v4}
\end{fmfgraph*}}}

\usepackage{tikz}
 \usetikzlibrary{calc,decorations.pathreplacing}

\def\Ree{\Re\textrm{e}}
\def\Imm{\Im\textrm{m}}

\def\be(#1,#2){{e^{i\pi \alpha' k_{#1}\cdot k_{#2}}}}
\def\bem(#1,#2){{e^{-i\pi \alpha' k_{#1}\cdot k_{#2}}}}
\def\trF{\textrm{Tr}}
\def\mS{\mathfrak S}

\def\cA{\mathcal A}

\newcount\hh
\newcount\mm
\mm=\time
\hh=\time
\divide\hh by 60
\divide\mm by 60
\multiply\mm by 60
\mm=-\mm
\advance\mm by \time
\def\thistime{\number\hh:\ifnum\mm<10{}0\fi\number\mm}

\begin{document}
\preprint{DAMTP-2016-53, IPhT-t16/069}

\title{Higher-loop amplitude  monodromy relations in string and gauge theory}

\author{Piotr Tourkine$^\dagger$, Pierre Vanhove$^{\dagger,\ddagger}$ \vspace{.1cm}
\\ \small{$^\dagger$DAMTP, University of Cambridge, Wilberforce Road, Cambridge CB3 0WA, UK
\\ $^\ddagger$Institut de physique th\'eorique, Universit\'e Paris Saclay, CNRS, F-91191 Gif-sur-Yvette, France
}}
\begin{abstract}
The monodromy relations in string theory provide a powerful and
elegant formalism to understand some of the deepest properties of
tree-level field theory amplitudes, like the color-kinematics duality.
This duality has been instrumental in tremendous progress on the
computations of loop amplitudes in quantum field theory, but a
higher-loop generalisation of the monodromy construction was lacking.

In this letter, we extend the monodromy relations to higher loops in
open string theory. Our construction, based on a contour deformation
argument of the open string diagram integrands, leads to new identities that
relate planar and non-planar topologies in string theory. We write one
and two-loop monodromy formul\ae{} explicitly at any multiplicity. In
the field theory limit, at one-loop we obtain identities that
reproduce known results.  At two loops, we check our formul\ae{} by
unitarity in the case of the four-point $\mathcal{N}=4$
super-Yang-Mills amplitude.
\end{abstract}

\maketitle
The search for the fundamental properties of the interactions between
elementary particles has been the driving force to uncover basic and
profound properties of scattering amplitudes in quantum field theory
and string theory.  In particular, the colour-kinematic
duality~\cite{Bern:2008qj} has led to tremendous progress in the
evaluation of loop amplitudes in gauge
theories~\cite{Bern:2010ue,Bern:2010tq,Boels:2012ew,Mafra:2014gja,Nohle:2013bfa,Bern:2013yya,Badger:2015lda,Mafra:2011kj,Mafra:2015mja,Chester:2016ojq,Primo:2016omk,Geyer:2015bja,Carrasco:2011hw}.
One remarkable consequence of this duality is the 
discovery of unsuspected kinematic relations between tree-level gauge
theory amplitudes~\cite{Bern:2008qj}, generated by a few fundamental
relations~\cite{BjerrumBohr:2009rd,Stieberger:2009hq,Feng:2010my,Johansson:2015oia,delaCruz:2015dpa}.

The monodromies of the open string disc
amplitudes~\cite{BjerrumBohr:2009rd,Stieberger:2009hq} did provide a
rationale for the kinematic relations between amplitudes at tree-level
in gauge theory. However, while the colour-kinematics duality has been
successfully implemented up to the fourth loop order in field
theory~\cite{Bern:2010tq,Boels:2012ew}, there is not yet a systematic understanding
of its validity to all loop orders.  It is therefore natural to seek a
higher-loop generalisation of the string theory approach to these
kinematic relations.

In this paper we generalise the tree-level monodromy construction to higher-loop open string diagrams (worldsheets with holes). This allows us derive new
relation between planar and non-planar topologies of
graphs in string theory.
The key ingredient in the construction relies on using a
representation of the string integrand with a loop momentum integration.
This is crucially needed in order to be able to understand zero mode shifts
when an external state jumps from one boundary to another.
Furthermore, just like at tree-level, the
construction does not depend on the precise nature of the scattering
amplitude nor the type of theory (bosonic or supersymmetric)
considered.

The relations that we obtain in field theory emanate from the leading
and first order in the expansion in the inverse string tension
$\alpha'$. At leading order, we find identities between planar and
non-planar amplitudes. At the next order, stringy corrections vanish
and we find the loop monodromy relations. They are relations between
integrands up to total derivatives, that involve both loop and
external momenta. Upon integration, this give relations between
amplitude-like integrals with extra powers of loop momentum in the
numerator.

At one loop, our string theoretic construction reproduces the field
theory relations of~\cite{Boels:2011tp,Boels:2011mn,Du:2012mt}. In
observing how the loop momentum factors produce cancellations of
internal propagators, we see that BCJ colour-kinematic representations
for numerators~\cite{Bern:2008qj} satisfy the monodromy relation at
the integrand level. The generality of our construction lead us to
conjecture that our monodromies generate all the kinematic relations
at any loop order.

We conclude by showing how our construction extends to higher
loops in string theory. In particular we write the two-loop string
monodromy relations. The field theory limit is subtle to understand in the
general case, but we provide a proof of concept with an example in $\mathcal{N}=4$
super-Yang-Mills at four-point two-loop, which we check by unitarity. 
We leave the general field theory relations for future work.

\vspace{-.5cm}
\section{Monodromies on the annulus}

One-loop  $n$-particle   amplitudes $\mathfrak A$ in oriented open-string
theory are defined on the annulus. They have a $U(N)$ gauge group and
the following colour decomposition~\cite{Green:1987mn}
\vspace{-10pt}
\begin{multline}\label{e:Atotal}
\mathfrak A(\{\epsilon_i,k_i,a_i\}  )= g_s^n
  \pi^{n-1}\sum_{p=0}^{n}\,\,\sum_{\alpha\cup\beta\in
  \mathfrak S_{p,n}}\cr
\trF(\lambda^{a_{\alpha(1)}}\cdots \lambda^{a_{\alpha(p)}}) \,
\trF(\lambda^{a_{\beta(p+1)}}\cdots \lambda^{a_{\beta(n)}}) \,
 \cA(\alpha|\beta)\,.
\end{multline}
The summation over $\mS_{p,n}$ of the external states distributed on
the boundaries of the annulus consists of permutations modulo cyclic
reordering and reflection symmetry. The quantities $k_i,\epsilon_i$
and $\lambda^a$ are the external momenta, polarizations
and colour matrices in the $U(N)$ fundamental representation, respectively. 
Planar amplitudes are obtained for $p=0$ or $p=n$ with $\trF(1)=N$.
The color-stripped ordered $n$-gluon amplitude $\cA(\alpha|\beta)$
take the following generic form  in $D$ dimensions
\begin{multline}\label{e:cAdef}
  \cA(\alpha|\beta)= \int_0^\infty \!\!\!dt \int_{\Delta_{\alpha|\beta}}  \hspace{-15pt}
  d^{n-1}\nu   \int
d^{D}\ell \, e^{-\pi   \alpha' t \ell^2-2i \pi \alpha'  \ell\cdot
  \sum_{k=1}^n k_i\nu_i}\\
\prod_{1\leq r<s\leq n}\!\!\!
f(e^{-2\pi t},\nu_r-\nu_s)
\times e^{-\alpha' k_r\cdot k_s\,
  G(\nu_r,\nu_s)}
 \,,
\end{multline}
where $t\in \mathbb{R}$ is the modulus of the annulus and the $\nu_i$'s are the location of the gluons insertions on the string
worldsheet -- one of them is set to $it$ by translation invariance. The loop momentum $\ell^\mu$ is
defined as the average of the string momentum $ \partial X^\mu$~\cite{D'Hoker:1988ta};
\begin{equation}
  \ell^\mu = \int_0^{\frac12}
  d\nu\, {\partial X^\mu(\nu)\over \partial \nu} \,.
\end{equation}
The domain of integration $\Delta_{\alpha|\beta}$ is the union of the
ordered sets $\{ \Imm(\nu_{\alpha(1)})<\cdots<\Imm(\nu_{\alpha(p)})\}$ for
$\Ree(\nu_i)=0$ 
and $\{ \Imm(\nu_{\beta(p+1)})>\cdots>\Imm(\nu_{\beta(n)})\}$ for $\Ree(\nu_i)=\frac12$.

We will show that the kinematical relations at one-loop
arise exclusively from 
shifts in  the loop-momentum-dependent part and 
monodromy properties of the non-zero mode part of the Green's function
in~\eqref{e:cAdef}
\begin{equation}
   G(\nu_r,\nu_s)=    -\log {\vartheta_1 (\nu_r-\nu_s|it)\over \vartheta_1'(0)}\,.
\label{eq:G-def}
\end{equation}
We refer to the appendix for some properties of the
propagators between the same and different boundaries.

The function $f(e^{-2\pi t},\nu_r-\nu_s)$ contains all the
theory-dependence of the amplitudes. The crucial point of our analysis
is that\textit{ it does not have any monodromy, therefore the
  relations that we obtain are fully generic}. This function is a
product of partition functions, internal momentum lattice of
compactification to $D$ dimensions, and a prescribed polarisation
dependence~\cite{Green:1987mn,Green:1981ya,Mafra:2012kh,He:2015wgf}. The
latter is composed of derivatives of the Green's function. None of
these objects have monodromies: that is why the precise form of
$f$ does not matter for our analysis. This property carries over to
higher-loop orders.

\vspace{-.5cm}
\subsection{Local and global monodromies}
\label{sec:contour-deformation}

\begin{figure}[t]
\definecolor{cff0000}{RGB}{255,0,0}
\begin{tikzpicture}[y=0.80pt, x=0.80pt, yscale=-1.000000, xscale=1.000000, inner sep=0pt, outer sep=0pt]
  \path[draw=black,line join=miter,line cap=butt,even odd rule,line width=0.577pt]
    (398.8321,47.7061) -- (398.8321,92.7646) -- (238.6762,92.7646);
  \path[xscale=1.000,yscale=-1.000,fill=black,line join=miter,line cap=butt,line
    width=0.800pt] (227.5552,-46.1983) node[above right] (text4496) {$0$};
  \path[fill=black,line join=miter,line cap=butt,line width=0.800pt]
    (218.0565,98.6025) node[above right] (text4500) {$\frac12$};
  \path[xscale=0.709,yscale=1.410,fill=black,line join=miter,line cap=butt,line
    width=0.800pt] (568.1055,72.0646) node[above right] (text4504) {${it}+\frac12$};
  \path[draw=black,line join=miter,line cap=butt,even odd rule,line width=0.577pt]
    (310.2795,43.9512) -- (317.7893,51.4610);
  \path[draw=black,line join=miter,line cap=butt,even odd rule,line width=0.577pt]
    (317.7892,43.9512) -- (310.2795,51.4610);
  \path[draw=black,line join=miter,line cap=butt,even odd rule,line width=0.577pt]
    (276.2249,43.9512) -- (283.7347,51.4610);
  \path[draw=black,line join=miter,line cap=butt,even odd rule,line width=0.577pt]
    (283.7347,43.9512) -- (276.2249,51.4610);
  \path[fill=black,line join=miter,line cap=butt,line width=0.800pt]
    (276.8388,38.8735) node[above right] (text4516) {$\nu_2$};
  \path[fill=black,line join=miter,line cap=butt,line width=0.800pt]
    (310.6843,38.9549) node[above right] (text4520) {$\nu_3$};
  \path[fill=black,line join=miter,line cap=butt,line width=0.800pt]
    (396.3936,38.9219) node[above right] (text4524) {$\nu_p$};
  \path[draw=cff0000,line join=miter,line cap=butt,miter limit=4.00,even odd
    rule,line width=0.578pt] (241.9932,90.5513) -- (241.9932,50.7931) --
    (273.3207,50.7372) .. controls (273.3207,50.7372) and (274.5108,55.2437) ..
    (280.0938,55.2437) .. controls (285.6768,55.2437) and (286.2416,50.7103) ..
    (287.0625,50.7103) -- (307.4958,50.6580) .. controls (307.4958,50.6580) and
    (307.8016,55.2434) .. (313.9536,55.2434) .. controls (320.1056,55.2434) and
    (320.8759,50.6334) .. (320.8759,50.6334) --
    (334.8571,50.6200)(352.9642,50.6110) -- (361.3393,50.6060) .. controls
    (361.9328,53.0299) and (363.7548,54.4404) .. (367.2601,54.4426) .. controls
    (370.4708,54.4446) and (372.5356,53.4539) .. (373.0917,50.6010) --
    (391.5145,50.5893) .. controls (391.5145,50.5893) and (391.1179,52.9128) ..
    (392.4831,54.3732) .. controls (393.8484,55.8336) and (396.1733,55.3311) ..
    (396.1733,55.3311) -- (395.9690,90.1309) -- (372.8571,90.1309) .. controls
    (371.0924,87.2526) and (370.7312,85.9975) .. (367.3661,86.0237) .. controls
    (364.2241,86.0482) and (363.3414,87.5496) .. (361.2500,90.1309) --
    (344.5714,90.1309)(321.2500,90.1309) -- (282.9297,90.1309) .. controls
    (281.4749,86.3464) and (281.0560,86.0271) .. (275.6061,86.0271) .. controls
    (270.4082,86.0271) and (268.8322,86.5550) .. (268.2194,90.1309) --
    (241.5412,90.1309);
  \path[draw=cff0000,fill=cff0000,even odd rule,line width=0.289pt]
    (301.0867,90.1050) -- (302.6017,91.5938) -- (297.3320,90.1378) --
    (302.5755,88.5900) -- (301.0867,90.1050) -- cycle;
  \path[draw=cff0000,fill=cff0000,even odd rule,line width=0.289pt]
    (299.3695,50.6963) -- (297.8545,52.1850) -- (303.1243,50.7290) --
    (297.8807,49.1812) -- (299.3695,50.6963) -- cycle;
  \path[draw=cff0000,fill=cff0000,even odd rule,line width=0.289pt]
    (382.0001,50.6784) -- (380.4851,52.1672) -- (385.7549,50.7112) --
    (380.5114,49.1634) -- (382.0001,50.6784) -- cycle;
  \path[draw=cff0000,fill=cff0000,even odd rule,line width=0.289pt]
    (257.7420,50.6963) -- (256.2270,52.1850) -- (261.4967,50.7290) --
    (256.2532,49.1812) -- (257.7420,50.6963) -- cycle;
  \path[draw=cff0000,fill=cff0000,even odd rule,line width=0.289pt]
    (241.9892,64.6157) -- (240.5004,66.1307) -- (241.9564,60.8610) --
    (243.5042,66.1045) -- (241.9892,64.6157) -- cycle;
  \path[draw=black,fill=black,even odd rule,line width=0.289pt] (418.0559,47.6767)
    -- (416.5409,49.1655) -- (421.8107,47.7095) -- (416.5672,46.1617) --
    (418.0559,47.6767) -- cycle;
  \path[draw=black,fill=black,even odd rule,line width=0.289pt]
    (238.6576,107.1722) -- (237.1688,105.6572) -- (238.6248,110.9269) --
    (240.1726,105.6834) -- (238.6576,107.1722) -- cycle;
  \path[draw=black,line join=miter,line cap=butt,miter limit=4.00,even odd
    rule,line width=0.578pt] (238.8156,107.3529) -- (238.8156,47.5923) --
    (418.4456,47.5923);
  \path[draw=black,line join=miter,line cap=butt,even odd rule,line width=0.577pt]
    (242.4311,70.2353) -- (234.9214,66.4805);
  \path[draw=black,line join=miter,line cap=butt,even odd rule,line width=0.577pt]
    (242.4311,73.2392) -- (234.9214,69.4844);
  \path[draw=black,line join=miter,line cap=butt,even odd rule,line width=0.577pt]
    (402.5870,70.2353) -- (395.0772,66.4805);
  \path[draw=black,line join=miter,line cap=butt,even odd rule,line width=0.577pt]
    (402.5870,73.2392) -- (395.0772,69.4844);
  \path[draw=cff0000,fill=cff0000,even odd rule,line width=0.289pt]
    (396.0003,74.5133) -- (394.5116,72.9983) -- (395.9675,78.2681) --
    (397.5153,73.0246) -- (396.0003,74.5133) -- cycle;
  \path[draw=black,line join=miter,line cap=butt,even odd rule,line width=0.577pt]
    (395.0981,43.9512) -- (402.6078,51.4610);
  \path[draw=black,line join=miter,line cap=butt,even odd rule,line width=0.577pt]
    (402.6078,43.9512) -- (395.0981,51.4610);
  \path[fill=black,line join=miter,line cap=butt,line width=0.800pt]
    (428.0358,50.6736) node[above right] (text4560) {$i\mathbb{R}$};
  \path[xscale=0.507,yscale=1.973,fill=black,line join=miter,line cap=butt,line
    width=0.800pt] (484.7275,34.3692) node[above right] (text4564)
    {${\color{red}\mathcal C}$};
  \path[draw=cff0000,fill=cff0000,dash pattern=on 1.73pt off 1.73pt,line
    join=miter,line cap=butt,miter limit=4.00,even odd rule,line width=0.578pt]
    (336.2591,50.6268) -- (352.2946,50.6268);
  \path[draw=black,line join=miter,line cap=butt,even odd rule,line width=0.577pt]
    (363.5791,43.9512) -- (371.0889,51.4610);
  \path[draw=black,line join=miter,line cap=butt,even odd rule,line width=0.577pt]
    (371.0889,43.9512) -- (363.5791,51.4610);
  \path[xscale=0.649,yscale=1.541,fill=black,line join=miter,line cap=butt,line
    width=0.800pt] (555.4528,25.2536) node[above right] (text4591) {$\nu_{p-1}$};
  \path[draw=black,line join=miter,line cap=butt,even odd rule,line width=0.577pt]
    (363.5791,89.0361) -- (371.0889,96.5458);
  \path[draw=black,line join=miter,line cap=butt,even odd rule,line width=0.577pt]
    (371.0889,89.0361) -- (363.5791,96.5458);
  \path[xscale=0.649,yscale=1.541,fill=black,line join=miter,line cap=butt,line
    width=0.800pt] (416.7413,68.8016) node[above right] (text4614) {$\nu_{n}$};
  \path[draw=cff0000,fill=cff0000,dash pattern=on 1.73pt off 1.73pt,line
    join=miter,line cap=butt,miter limit=4.00,even odd rule,line width=0.578pt]
    (320.4016,90.1301) -- (343.2557,90.1301);
  \path[draw=black,line join=miter,line cap=butt,even odd rule,line width=0.577pt]
    (271.5791,89.0361) -- (279.0889,96.5458);
  \path[draw=black,line join=miter,line cap=butt,even odd rule,line width=0.577pt]
    (279.0889,89.0361) -- (271.5791,96.5458);
  \path[xscale=0.649,yscale=1.541,fill=black,line join=miter,line cap=butt,line
    width=0.800pt] (549.9360,68.8016) node[above right] (text4637) {$\nu_{p+1}$};
  \path[draw=cff0000,fill=cff0000,even odd rule,line width=0.289pt]
    (344.0001,50.6784) -- (342.4851,52.1672) -- (347.7549,50.7112) --
    (342.5114,49.1634) -- (344.0001,50.6784) -- cycle;
  \path[xscale=0.507,yscale=-1.973,fill=black,line join=miter,line cap=butt,line
    width=0.800pt] (584.4955,-41.0925) node[above right] (text4564-2)
    {${\color{red}\nu_1}$};
  \path[xscale=1.000,yscale=-1.000,fill=black,line join=miter,line cap=butt,line
    width=0.800pt] (466.4286,-87.7250) node[above right] (text5089) {$$};
  \path[draw=black,fill=black,even odd rule,line width=0.289pt] (289.7872,47.6769)
    -- (288.2722,49.1657) -- (293.5420,47.7097) -- (288.2985,46.1619) --
    (289.7872,47.6769) -- cycle;
  \path[draw=black,fill=black,even odd rule,line width=0.289pt] (331.7872,47.6769)
    -- (330.2722,49.1657) -- (335.5420,47.7097) -- (330.2985,46.1619) --
    (331.7872,47.6769) -- cycle;
  \path[draw=black,fill=black,even odd rule,line width=0.289pt] (327.9210,92.7305)
    -- (329.4360,94.2193) -- (324.1663,92.7633) -- (329.4098,91.2155) --
    (327.9210,92.7305) -- cycle;
  \path[fill=black,line join=miter,line cap=butt,line width=0.800pt]
    (242.4366,115.4457) node[above right] (text4223) {$\mathbb{R}$};
\end{tikzpicture}
\caption{The $\nu_1$ contour integral (red) vanishes. The two
  boundaries (black) have opposite orientation.}
   \label{fig:FFT}
 \end{figure}
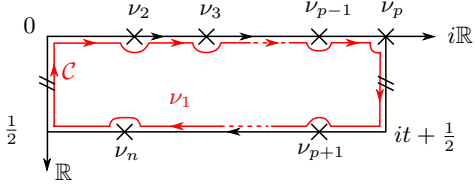

 Let us consider the non-planar amplitude $\mathcal{A}(1,\ldots,p |
 p+1,\ldots,n)$, but where we take the modified integration contour
 $\mathcal{C}$ of fig.~\ref{fig:FFT} for
 $\nu_1$. The integrand being
 holomorphic, in virtue of Cauchy's theorem, the integral vanishes: %
\begin{multline}
  \oint_{\mathcal C}  d\nu_1 \int_0^\infty \!\!\!d^{D}\ell \, e^{-\pi
    \alpha' t \ell^2-2 i\pi \alpha' \ell\cdot \sum_{k=2}^n
    k_i\nu_i}e^{-2i\pi\alpha' \ell \cdot k_1 \, \nu_1}  \times\cr
  \prod_{r=2}^n\,f(e^{-2\pi t},\nu_1-\nu_r)
e^{-\alpha'\, k_1\cdot k_r\,  G(\nu_1,\nu_r)} =0  \,.
\end{multline}
Each separate portion of the integration corresponds to a different
ordering and topology. The portions along the vertical sides cancel by
periodicity of the one-loop integral (cf.  appendix). We are thus left
with the contributions from the boundaries
$\Ree(\nu_1)=0$ and $\Ree(\nu_1)=\frac12$.
When exchanging the position of two states 
on the \emph{same} boundary, the short
distance behaviour of the Green's function $G(\nu_1,\nu_2)\simeq
-\log (\nu_1-\nu_2)$ implies
  \begin{equation}
 G(\nu_1,\nu_2)= G(\nu_2,\nu_1)\pm i\pi\,,
  \end{equation}
with $-i\pi$ for a clockwise rotation and $+i\pi$ for a
counter-clockwise rotation.
Thus, on the upper part of the contour in figure~\ref{fig:FFT}, exchanging the positions of two external states leads to an phase factor multiplying the amplitude
\begin{equation}
  \cA(12\cdots m| m+1\cdots n)\to  e^{i\pi \alpha' k_1\cdot k_2}  \,
  \cA(21\cdots m| m+1\cdots  n)  
\end{equation}
On the lower part of the contour in figure~\ref{fig:FFT}, the phases come with the same sign due to an
additional sign from $\vartheta_2$ in eq.~\eqref{eq:theta1-half-period}. 
For external states on different boundaries, the Green's
function involves the even function $\vartheta_2(\nu_r-\nu_s)$ and the ordering does not
matter (cf. the appendix).

The main difference with the tree-level case arises from the
\emph{global} monodromy transformation when a state 
moves from one boundary to the other, $\nu_1\to \nu_1+\frac12$. This
produces a new phase $\exp(- i\pi \alpha' \ell \cdot k_1)$ in the
integrand
\begin{multline}\label{e:ANPtilde}
\cA(12\cdots n)\to\cA(2\cdots n|1)[e^{-i\pi \alpha' \ell\cdot k_1}]:=\cr\int_0^\infty dt\int_{\Delta_{2\cdots n|1}}  \hspace{-20pt} d^{n-1}\nu \, \prod_{1\leq
  r<s\leq n} f(e^{-2\pi t},\nu_r-\nu_s) \,
 e^{-\alpha' k_r\cdot k_s\,    G(\nu_r,\nu_s)}\cr
\times\int_0^\infty d^{D}\ell \, e^{-  i\pi \alpha'
    \ell \cdot k_1}  \, e^{-\pi     \alpha' t \ell^2-2i \pi \alpha' \ell\cdot \sum_{k=1}^n k_i\nu_i}
\,.
\end{multline}

On non-orientable surfaces the
propagator is obtained by appropriate shifts of the Green's function~\eqref{eq:G-def} according the effects of the twist
operators~\cite{Green:1981ya}.  The local monodromies are the same
because they only depend on the short distance behaviour of the
propagator, and global monodromies are obtained  in an immediate generalisation of our
construction.

\subsection{Open string relations}

We can now collect up all the previous pieces. Paying great care to
signs and orientations, according to what was described, the vanishing
of the integral along $\mathcal{C}$ gives the following generic
relation\footnote{Compared to earlier versions, we correct here a sign
  mistake in the non-planar phases. Because of this mistake, in
  fig.~\ref{fig:FFT}, we took the cuts of the non-planar vertical
  $\Ree (\nu_1)=1/2$ contour to be downard cuts, the corrected version
  has upward cuts. The analysis for the $\Ree (\nu_1)=0$ cuts is
  unchanged. Details on the correct version are given
  in~\cite[Appendix~B]{Ochirov:2017jby}.}
\begin{multline}
  \label{eq:monod-n-generic}
  \cA(1,2,\ldots ,p|p+1, \ldots ,n) 
 +\cr  
\sum_{i=2}^{p-1} e^{i\alpha'\pi k_1\cdot k_{2\cdots i} }
  \cA(2,\ldots ,i,1,i+1,\ldots,p |p+1,\ldots ,n)=\cr
 -\sum_{i=p}^n \big(  e^{-i\alpha'\pi k_1\cdot k_{i+1\cdots n}}\times\cr
  \cA(2,\ldots ,p|p+1 ,\ldots ,i,1,i+1,\ldots ,n)
  [e^{-i\pi\alpha'\ell\cdot k_1}]\big)
\end{multline}
where the bracket notation was defined in~\eqref{e:ANPtilde} and we
set $k_{1\ldots p}:=\sum_{i=1}^pk_i$.  In particular, starting from the
planar four-point amplitude we find the following formula
\begin{multline}
\cA(1234)+ e^{i\pi\alpha' k_1\cdot k_2} \cA(2134)+e^{i\pi\alpha'
  k_1\cdot (k_2+k_3)}\cA(2314)=\cr -\cA(234|1)[e^{-i\pi\alpha'\ell\cdot k_1}]\,.
\label{eq:monod-4pt}
\end{multline}
We also find, starting from a purely planar amplitude 
\begin{multline}
  \label{e:multipleflip}
(-1)^{|\beta|}\sum_{\gamma\in\alpha\shuffle\beta}  \prod_{a=1}^s\prod_{b=1}^r e^{i\pi\alpha'(\alpha_a,\beta_b)}\, \cA(\gamma_1\cdots \gamma_{r+s}\,n)=\cr
\cA(\alpha_1\cdots\alpha_s\,n| \beta_r\cdots
\beta_1)\big[\prod_{i=1}^r e^{-i\pi\alpha'\ell\cdot k_{\beta_i}}\big]
\end{multline}
where now we integrate
the vertex operators with ordered position $\Imm(\nu_{\beta_1})\leq\cdots
\leq\Imm(\nu_{\beta_r})$
along the contour of fig.~\ref{fig:FFT}.
The sum is over the shuffle product  $\alpha\shuffle\beta$ and 
 the permutation $\beta$ of length $|\beta|$, and $(\alpha_i,\beta_j)=k_{\alpha_i}\cdot
  k_{\beta_j}$ if $\Imm(\nu_{\beta_j})>\Imm(\nu_{\alpha_i})$ in $\gamma$ and 1 otherwise.
The phase factors with external momenta are the same as at
tree-level: the new ingredients here are the insertions of loop-momentum dependent
factors inside the integral.

Note that some of our relations involve objects like
$\cA(2\cdots n|1)$ that seemingly contribute in~\eqref{e:Atotal} only
if the state 1 is a colour singlet. However, our relations involve
colour-stripped objects and are, therefore, valid in full generality.
Note also that our relations are valid under the $t$-integration,
thus they are not affected by the dilaton tadpole divergence at
$t\to0$~\cite{Green:1981ya}.

We have thus shown that the kinematic
relations~\eqref{eq:monod-n-generic} relate planar and non-planar open
string topologies, which normally have independent colour
structures.  This is the one-loop generalisation of the string theory fundamental
monodromies that generates all amplitude relations at tree-level in
string theory~\cite{BjerrumBohr:2009rd,Stieberger:2009hq}.  Thus, we
conjecture our one-loop relations~\eqref{eq:monod-n-generic},
written for all the permutations of the external states, generate all
the one-loop oriented open string theory relations. Let us now turn to
the consequences in field theory.

\section{Field theory relations}
\label{sec:field-theory-limit-1}
Gauge theory amplitudes are extracted from string theory ones in the standard way.
We send $\alpha'\to0$ and keep fixed the quantity $\alpha' t$ that becomes the Schwinger
proper-time in field theory. We also set $\Imm(\nu)=x\, t$, with $0\leq x\leq1$.
The Green's function of eq.~\eqref{eq:G-def} reduces to the sum of the field
theory worldline propagator $x^2 - |x|$ and a stringy correction
\begin{equation}\label{e:Gdelta}
 G(\nu)=  t\,  \left(x^2 -
   |x|\right)+\delta_{\pm}(x)+O(e^{-2\pi t})\,.
\end{equation}
(for details see appendix).\footnote{In bosonic open
  string one would need to keep to the terms of the order
  $\exp(-2\pi t)$ because of the Tachyon.} At leading order in
$\alpha'$, open string amplitudes reduce to the 
usual parametric
representation of the dimensional regulated gauge theory amplitudes~\cite{Green:1982sw,Bern:1992cz}.\footnote{See also
  \cite{Bern:1990cu,Bern:1990ux,Bern:1991aq} for equivalent closed
  string methods} All the
monodromy phase factors reduce to $1$ and
from~\eqref{e:multipleflip} we recover the well-known photon
decoupling relations between non-planar and planar
amplitudes~\cite{Bern:1994zx}, with
$\beta^T=(\beta_r,\ldots,\beta_1)$,
\begin{equation}\label{e:ANPP}
A(\alpha|\beta^T)=(-1)^{|\beta|} \sum_{\gamma\in \alpha\shuffle\beta} \,A(\gamma)\,.
\end{equation}
This is an important consistency check on our relations.

At the first order in $\alpha'$ we get contributions from expansion of
the phase factors but as well potential ones from the massive stringy
mode coming from $\delta_\pm(x)$.  The analysis of the appendix
of~\cite{Green:1999pv} shows that this contributes to next order in
$\alpha'$, which, importantly, allow us to neglect it here.
Therefore, the field theory limit of~\eqref{eq:monod-n-generic} gives a new identity
\begin{multline}
  \label{eq:monod-n-genericQFT}
\sum_{i=2}^{p-1} k_1\cdot   k_{2\cdots i}\,A(2,\ldots, i, 1, i+1,\ldots ,p|p+1,\ldots ,n) +\cr 
\sum_{i=p}^{n}    k_1\cdot k_{i+1\cdots n}
  \,A(2,\ldots ,p|p+1,\ldots ,i ,1, i+1, \ldots, n )=\cr
  \sum_{i=p}^{n} \,A(2,\ldots ,p|p+1, \ldots ,i,1,i+1,\ldots, n) [\ell\cdot k_1]\,.
\end{multline}
These relations  are the one-loop equivalent of the fundamental
monodromy identities~\cite{Feng:2010my,Johansson:2015oia,delaCruz:2015dpa} that
generates all the amplitude relations at tree-level.

In particular, using~\eqref{e:ANPP}, we obtain the relation between
planar gauge theory integrands with  linear power of loop momentum 
  \begin{multline}
   A(1\cdots n)[ \ell\cdot k_1]+
A(21\cdots n)[(\ell+k_2)\cdot k_1 ]+\cdots +\\
A(23\cdots (n-1) 1n)[(\ell+k_{23\ldots {n-1}})\cdot
k_1]=0\,.
\label{eq:n-pt-planar}
  \end{multline}
These are the relations derived
in~\cite{Boels:2011tp,Boels:2011mn,Du:2012mt}: this
constitutes an additional check on our formul\ae.

Let us now analyse the effect of the linear momentum factors at the level
of the graphs. 
At this point we pick any representation of the integrand in terms of
cubic graphs only and the field theory limit defines the loop momentum
as the internal momentum following immediately the leg
$n$.\footnote{This is checked by matching with usual definition of the  Schwinger proper times.}  We then
rewrite the loop momentum factors as differences of
propagators. Hence, each individual graph with numerator $n_G$
produces two graphs with one fewer propagator, e.g.
 \begin{fmffile}{pent-red}
\vspace{4pt}
  \begin{equation}
    \ell\cdot k_1 \begin{gathered}
{\pentg{1}{2}{3}{4}{5}}
\end{gathered}\quad=\quad
\begin{gathered}
  \parbox{35pt}{
\begin{fmfgraph*}(32,26) 
\fmfleft{i1,i2}
\fmfright{i0,i3} 
\fmf{phantom}{i1,v1}
\fmf{plain}{i2,v2}
\fmf{plain}{i3,v3}
\fmf{plain}{i0,v4}
\fmfv{label=${\color{blue}3}$,label.dist=3pt}{i2}
\fmfv{label=${\color{blue}4}$,label.dist=3pt}{i3}
\fmfv{label=${\color{blue}5}$,label.dist=3pt}{i0}
\fmf{plain,tension=0.5}{v1,v2,v3,v4,v1} 
\fmf{fermion,tension=0,label=$\ell$,label.side=left}{v4,v1}
\fmffreeze
\fmfipair{xx,xy} 
\fmfiequ{xx}{(0.25w,-.1w)} 
\fmfiequ{xy}{(+0.h,0.2h)} 
 \fmfiv{l=${\color{blue}1}$,l.a=-115,l.d=3pt}{xx}
 \fmfiv{l=${\color{blue}2}$,l.a=180,l.d=3pt}{xy}
\fmfcmd{draw xx--vloc(__v1); draw xy--vloc(__v1);}
\end{fmfgraph*}}
\end{gathered}\quad-\quad
\begin{gathered}
  \parbox{35pt}{
\begin{fmfgraph*}(32,26) 
\fmfleft{i1,i2}
\fmfright{i0,i3} 
\fmf{plain}{i1,v1}
\fmf{plain}{i2,v2}
\fmf{plain}{i3,v3}
\fmfv{label=${\color{blue}3}$,label.dist=3pt}{i2}
\fmfv{label=${\color{blue}4}$,label.dist=3pt}{i3}
\fmfv{label=${\color{blue}2}$,label.dist=3pt}{i1}
\fmf{phantom}{i0,v4}
\fmf{plain,tension=0.5}{v1,v2,v3,v4,v1} 
\fmf{fermion,tension=0,label=$\ell\!-\!k_5$,label.side=left,label.dist=8pt}{v3,v4}
\fmffreeze
\fmfipair{xx,xy} 
\fmfiequ{xx}{(w-0.25w,-.1w)} 
\fmfiequ{xy}{(w+0.h,0.2h)} 
 \fmfiv{l=${\color{blue}1}$,l.a=-45,l.d=3pt}{xx}
 \fmfiv{l=${\color{blue}5}$,l.a=0,l.d=3pt}{xy}
\fmfcmd{draw xx--vloc(__v4); draw xy--vloc(__v4);}
\end{fmfgraph*}}
\end{gathered}
\label{eq:pent-red}
\end{equation}
\end{fmffile}
Then, there always exist another graph $G'$ that will produce one of
the two reduced graphs as well, with a different numerator
$n_{G'}$. In the previous example, it would be the $21345$ pentagon
for the massive box with $1,2$ corner. 
Finally, reduced graphs also arise directly from string theory, when
vertex operators collide~\cite{Bern:1992cz}. In
\eqref{eq:n-pt-planar}, these always appear in such combinations of two
graphs, say $G_1$ and $G_2$;
\begin{fmffile}{box-box}
 \begin{equation}
\label{eq:boxbox}
\ell\cdot k_1 \parbox[c][30pt][c]{30pt}{
 \begin{fmfgraph*}(26,20) 
\fmfleft{i1,i2}
\fmfright{i0,i3} 
\fmf{plain}{i1,v1}
\fmf{plain}{i2,v2}
\fmf{plain}{i3,v3}
\fmfv{label=${\color{blue}3}$,label.dist=3pt}{i2}
\fmfv{label=${\color{blue}4}$,label.dist=3pt}{i3}
\fmfv{label=${\color{blue}5}$,label.dist=3pt}{i0}
\fmf{plain}{i0,v4}
\fmf{plain,tension=0.5}{v1,v2,v3,v4,v1}
\fmffreeze
\fmfipair{xx,xy} 
\fmfiequ{xx}{(0.1w,-.38w)} 
\fmfiequ{xy}{(-0.38h,0)} 
 \fmfiv{l=${\color{blue}1}$,l.a=-90,l.d=3pt}{xx}
 \fmfiv{l=${\color{blue}2}$,l.a=180,l.d=3pt}{xy}
\fmfcmd{draw xx--vloc(__i1); draw xy--vloc(__i1);}
\end{fmfgraph*}}\quad+\quad (\ell+k_2)\cdot k_1
\parbox[c][30pt][c]{30pt}{
\begin{fmfgraph*}(26,20) 
\fmfleft{i1,i2}
\fmfright{i0,i3} 
\fmf{plain}{i1,v1}
\fmf{plain}{i2,v2}
\fmf{plain}{i3,v3}
\fmfv{label=${\color{blue}3}$,label.dist=3pt}{i2}
\fmfv{label=${\color{blue}4}$,label.dist=3pt}{i3}
\fmfv{label=${\color{blue}5}$,label.dist=3pt}{i0}
\fmf{plain}{i0,v4}
\fmf{plain,tension=0.5}{v1,v2,v3,v4,v1} 
\fmffreeze
\fmfipair{xx,xy} 
\fmfiequ{xx}{(0.1w,-.38w)} 
\fmfiequ{xy}{(-0.38h,0)} 
 \fmfiv{l=${\color{blue}2}$,l.a=-90,l.d=3pt}{xx}
 \fmfiv{l=${\color{blue}1}$,l.a=180,l.d=3pt}{xy}
\fmfcmd{draw xx--vloc(__i1); draw xy--vloc(__i1);}
\end{fmfgraph*}}
\vspace{7pt}
\end{equation}
\end{fmffile}

\noindent The color ordered 3-point vertex is antisymmetric, so $n_{G_1}=-n_{G_2}$ and the
$\ell\cdot k_1$ terms cancel.
We then realize that the graphs entering the monodromy relations can
be organised by triplets of Jacobi numerators $n_G+n_{G'}-n_{G_1}$
times denominator. In a BCJ representation, all these triplets vanish
identically and eq.~\eqref{eq:monod-n-genericQFT} is satisfied at the
integrand level. Thus, any BCJ representation satisfies these
monodromy relations, but the converse is not true.

\section{Toward  Higher-loop relations}
\label{sec:discussion}

\begin{figure}[h]
  \centering
 \definecolor{c808080}{RGB}{128,128,128}
\definecolor{cff0000}{RGB}{255,0,0}
\definecolor{c0000ff}{RGB}{0,0,255}

\begin{tikzpicture}[y=0.80pt, x=0.80pt, yscale=-1.000000, xscale=1.000000, inner sep=0pt, outer sep=0pt]
  \path[draw=c808080,dash pattern=on 1.60pt off 0.80pt,line join=miter,line
    cap=butt,miter limit=4.00,even odd rule,line width=0.400pt] (203.1017,28.8561)
    .. controls (193.1249,23.1588) and (175.9504,26.0678) .. (168.8010,37.4863) ..
    controls (161.6516,48.9047) and (165.0986,65.9800) .. (171.6974,72.6986) ..
    controls (178.2962,79.4171) and (178.9704,83.6206) .. (177.6659,89.6347);
  \path[draw=c808080,dash pattern=on 1.60pt off 0.80pt,line join=miter,line
    cap=butt,miter limit=4.00,even odd rule,line width=0.400pt] (210.9622,33.9174)
    .. controls (221.7107,41.9668) and (222.5856,58.3983) .. (213.3197,69.0959) ..
    controls (204.0539,79.7935) and (187.0713,79.2949) .. (177.7307,89.6904);
  \path[draw=cff0000,line join=miter,line cap=butt,miter limit=4.00,even odd
    rule,line width=0.578pt] (201.1315,32.0630) .. controls (194.4219,28.6817) and
    (186.2580,28.9294) .. (179.5828,33.2622) .. controls (169.2914,39.9423) and
    (166.4357,53.8029) .. (173.2044,64.2207) .. controls (179.9731,74.6385) and
    (193.8031,77.6685) .. (204.0945,70.9885) .. controls (214.3860,64.3084) and
    (217.2418,50.4478) .. (210.4730,40.0300) .. controls (209.8974,39.1441) and
    (209.2707,38.3116) .. (208.6180,37.5673) .. controls (211.6121,32.7860) and
    (215.3742,27.4168) .. (217.0242,19.2453) .. controls (218.9769,23.6979) and
    (223.7179,32.2640) .. (230.0645,37.2158) .. controls (228.6137,38.8430) and
    (227.5068,40.6863) .. (226.5485,42.7406) .. controls (221.2962,53.9986) and
    (226.0541,67.3270) .. (237.1755,72.5105) .. controls (248.2970,77.6940) and
    (261.5704,72.7696) .. (266.8227,61.5116) .. controls (272.0750,50.2536) and
    (267.3171,36.9251) .. (256.1957,31.7417) .. controls (249.8588,28.7882) and
    (242.8234,29.1163) .. (237.1436,31.9619) .. controls (231.4125,27.6857) and
    (223.2603,16.2463) .. (221.2954,11.2425) .. controls (223.3575,11.2907) and
    (222.9500,11.3167) .. (225.0105,11.3750) .. controls (241.5592,11.8432) and
    (256.9401,14.1611) .. (270.0516,17.8417) -- (276.0010,19.7252) .. controls
    (298.4337,27.1161) and (312.9251,38.8513) .. (312.9251,52.0521) .. controls
    (312.9251,74.5659) and (270.7734,92.8170) .. (218.7767,92.8170) .. controls
    (166.7800,92.8170) and (124.6283,74.5659) .. (124.6283,52.0521) .. controls
    (124.6283,38.0648) and (140.8980,25.7229) .. (165.5497,18.3660) --
    (167.9046,17.6914) -- (173.2804,16.4873) .. controls (183.7115,14.1083) and
    (195.7250,12.2739) .. (208.1944,11.5841) -- (211.4654,11.3746) .. controls
    (208.2004,20.9776) and (202.6119,30.1911) .. (201.1315,32.0630) -- cycle;
  \path[draw=c0000ff,line join=miter,line cap=butt,miter limit=4.00,even odd
    rule,line width=0.400pt] (203.3170,36.6588) .. controls (209.1058,27.0351) and
    (213.9252,20.3553) .. (216.3022,8.1954);
  \path[draw=c0000ff,line join=miter,line cap=butt,miter limit=4.00,even odd
    rule,line width=0.400pt] (235.7382,36.2333) .. controls (226.0217,28.5787) and
    (217.2953,15.7161) .. (216.2706,8.2510);
  \path[draw=black,miter limit=4.00,nonzero rule,line width=0.578pt]
    (218.7767,52.1253) ellipse (2.7837cm and 1.2435cm);
  \path[xscale=1.000,yscale=1.000,fill=black,line join=miter,line cap=butt,line
    width=0.800pt] (192.0175,23.0780) node[above right] (text4177) {${\color{blue}
    a_1}$};
  \path[draw=black,miter limit=4.00,nonzero rule,line width=0.578pt]
    (191.8382,52.1253) ellipse (0.5460cm and 0.5388cm);
  \path[draw=black,miter limit=4.00,nonzero rule,line width=0.578pt]
    (246.6948,52.1253) ellipse (0.5460cm and 0.5388cm);
  \begin{scope}[cm={{0.78134,0.0,0.0,0.78096,(63.01972,11.87214)}}]
    \path[draw=black,line join=miter,line cap=butt,miter limit=4.00,even odd
      rule,line width=0.739pt] (315.8018,56.6502) -- (323.9681,48.4839);
    \path[draw=black,line join=miter,line cap=butt,miter limit=4.00,even odd
      rule,line width=0.739pt] (323.9681,56.6502) -- (315.8018,48.4839);
  \end{scope}
  \path[draw=black,line join=miter,line cap=butt,miter limit=4.00,even odd
    rule,line width=0.578pt] (230.2801,11.9887) -- (236.6607,5.6112);
  \path[draw=black,line join=miter,line cap=butt,miter limit=4.00,even odd
    rule,line width=0.578pt] (236.6607,11.9887) -- (230.2801,5.6112);
  \path[draw=black,line join=miter,line cap=butt,miter limit=4.00,even odd
    rule,line width=0.578pt] (206.5252,48.2155) -- (212.9058,41.8380);
  \path[draw=black,line join=miter,line cap=butt,miter limit=4.00,even odd
    rule,line width=0.578pt] (212.9058,48.2155) -- (206.5252,41.8380);
  \path[draw=black,line join=miter,line cap=butt,miter limit=4.00,even odd
    rule,line width=0.578pt] (224.4026,59.4752) -- (230.7832,53.0977);
  \path[draw=black,line join=miter,line cap=butt,miter limit=4.00,even odd
    rule,line width=0.578pt] (230.7832,59.4752) -- (224.4026,53.0977);
  \path[draw=black,line join=miter,line cap=butt,miter limit=4.00,even odd
    rule,line width=0.578pt] (262.4158,52.3610) -- (268.7965,45.9834);
  \path[draw=black,line join=miter,line cap=butt,miter limit=4.00,even odd
    rule,line width=0.578pt] (268.7965,52.3610) -- (262.4158,45.9834);
  \path[draw=black,line join=miter,line cap=butt,miter limit=4.00,even odd
    rule,line width=0.578pt] (195.9947,12.4783) -- (202.3753,6.1007);
  \path[draw=black,line join=miter,line cap=butt,miter limit=4.00,even odd
    rule,line width=0.578pt] (202.3753,12.4783) -- (195.9947,6.1007);
  \path[draw=black,line join=miter,line cap=butt,miter limit=4.00,even odd
    rule,line width=0.578pt] (174.5309,92.8712) -- (180.9116,86.4937);
  \path[draw=black,line join=miter,line cap=butt,miter limit=4.00,even odd
    rule,line width=0.578pt] (180.9116,92.8712) -- (174.5309,86.4937);
  \path[draw=black,line join=miter,line cap=butt,miter limit=4.00,even odd
    rule,line width=0.578pt] (234.4433,98.3946) -- (240.8239,92.0170);
  \path[draw=black,line join=miter,line cap=butt,miter limit=4.00,even odd
    rule,line width=0.578pt] (240.8239,98.3946) -- (234.4433,92.0170);
  \path[xscale=1.000,yscale=1.000,fill=black,line join=miter,line cap=butt,line
    width=0.800pt] (231.8807,53.6650) node[above right] (text8735) {$$};
  \path[xscale=1.000,yscale=1.000,fill=black,line join=miter,line cap=butt,line
    width=0.800pt] (230.4939,23.8891) node[above right] (text8762) {${\color{blue}
    a_2}$};
  \path[draw=cff0000,fill=cff0000,even odd rule,line width=0.400pt]
    (266.8367,42.5117) -- (266.1119,45.2156) -- (264.3613,38.2263) --
    (269.5419,43.2362) -- (266.8367,42.5117) -- cycle;
  \path[draw=cff0000,fill=cff0000,line join=miter,even odd rule,line
    width=0.400pt] (208.2893,18.8988) -- (207.1471,15.6947) -- (205.6694,24.5455)
    -- (211.5275,17.5856) -- (208.2893,18.8988) -- cycle;
  \path[draw=black,fill=black,even odd rule,line width=0.400pt] (249.8272,10.2715)
    -- (247.5842,12.0482) -- (254.5439,11.0722) -- (248.2970,7.8540) --
    (249.8272,10.2715) -- cycle;
  \path[draw=black,fill=black,even odd rule,line width=0.400pt] (192.2046,33.3439)
    -- (194.1183,35.4710) -- (187.4203,33.3439) -- (194.1183,31.2167) --
    (192.2046,33.3439) -- cycle;
  \path[draw=black,fill=black,even odd rule,line width=0.400pt] (249.0676,33.0713)
    -- (250.9813,35.1984) -- (244.2834,33.0713) -- (250.9813,30.9442) --
    (249.0676,33.0713) -- cycle;
  \path[xscale=1.000,yscale=1.000,fill=black,line join=miter,line cap=butt,line
    width=0.800pt] (154.1256,67.8976) node[above right] (text6069) {$$};
  \path[xscale=1.000,yscale=1.000,fill=black,line join=miter,line cap=butt,line
    width=0.800pt] (172.1687,103.0161) node[above right] (text10509) {$P$};
  \path[xscale=1.000,yscale=1.000,fill=black,line join=miter,line cap=butt,line
    width=0.800pt] (152.7616,65.6166) node[above right] (text4177-9)
    {${\color{gray} \gamma_1}$};
  \path[xscale=1.000,yscale=1.000,fill=black,line join=miter,line cap=butt,line
    width=0.800pt] (198.0909,85.8989) node[above right] (text4177-9-3)
    {${\color{gray} \gamma_2}$};
  \path[draw=cff0000,fill=cff0000,line join=miter,even odd rule,line
    width=0.400pt] (212.1705,31.3749) -- (213.3128,34.5791) -- (214.7904,25.7283)
    -- (208.9324,32.6881) -- (212.1705,31.3749) -- cycle;

\end{tikzpicture}
  \caption{Two-loop integrand monodromy. Integration over the red
    contour vanishes. Given the definition of the loop momentum in
    eq.~\eqref{eq:higher-g-ell}, parallel integrations along $a_1,a_2$ 
  cancel only up to a shift in the loop momentum.
}
  \label{fig:two-loop}
\end{figure}
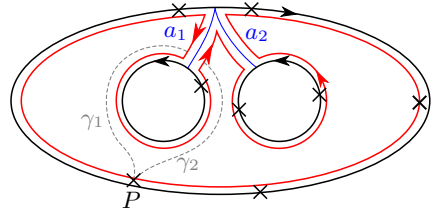

Higher-loop oriented open string diagrams are worldsheets with
holes, one for each loop.\footnote{
We do not consider string diagrams with handles in this work. They  lead to non-planar $1/N^2$
  corrections~\cite{Berkovits:2009aw}. 
} 
Just like at one loop, we consider the integral of the position of a
string state on a contractible closed contour that follows the
interior boundary of the diagram (cf. for instance
fig.~\ref{fig:two-loop}).  The integral vanishes without insertion of
closed string operator in the interior of the diagram. This
constitutes the essence of the monodromy relations at higher-loop.

Because the exchange of two external states on the \emph{same}
boundary depends only on the local behaviour of the Green's
function, we have the same \emph{local} monodromy transformation
$G(z_1,z_2)= G(z_2,z_1)\pm i\pi$ as at tree-level.  

Like at one loop, the \emph{global} monodromy of moving the external
state 1 from one boundary to another boundary by crossing the cycle
$a_I$ leads to the factor $\exp(-i\alpha'\pi \ell_I \cdot k_1)$.
The loop momenta $\ell_I$ are the zero-modes of the string momenta
$\ell_I=\int_{a_I} \partial X$~\cite{D'Hoker:1988ta}. The string
integrand depends on them through the factor:
\begin{equation}
  \label{eq:higher-g-ell}
\int \prod_{i=1}^g d\ell_i\, e^{\alpha' i\pi \sum_{I,J} \ell_I
  \ell_J \Omega_{IJ}-2i\pi \alpha' \sum_{I,j} \ell_I\cdot k_j
  \int^{z_j}_P \omega_I}\,,
\end{equation}
Importantly, the integration path between $P$ and $z_j$ in
\eqref{eq:higher-g-ell} depends on a homology class. This implies that
this expression has an intrinsic multivaluedness, corresponding to the
freedom of shifting the loop momentum by external momenta when
punctures cross through the $a$ cycles.\footnote{ Doing the Gaussian
  integration reduces to the standard expression of the string
  propagator, which is single valued on the surface.}. Choosing one
for each of these contours induces a choice of $g$ cuts on the
worldsheet along $g$ given $a$ cycles that renders the expression
single-valued. Our choice to make the $a$ cycle join at some common
point also removes the loop momentum shifting ambiguity and give
globally defined loop momenta.

%
 \paragraph{\bf A two-loop example.}
 \label{sec:two-loop-case}
 The generalisation of~\eqref{eq:monod-n-generic} gives the two-loop
 integrated relations\footnote{Compared to earlier versions, we
   corrected a sign in the non-planar phases. Higher-loop phases are
   related to the ones at one-loop by the factorisation limit of the
   string amplitude.}
\begin{multline}
    \label{e:mond-2-loop}
\sum_{r=1}^{|\alpha|} \Big(\prod_{s=1}^r e^{i \alpha'\pi k_1\cdot k_{\alpha_s}}\Big)
\cA^{(2)}(\ldots, \alpha_{s-1},1,\alpha_s,\ldots|\beta|\gamma)+\cr
\sum_{r=1}^{|\beta|} \Big(\prod_{s=1}^r e^{-i \alpha'\pi k_1\cdot k_{\beta_s}}\Big)
\cA^{(2)}(\alpha|\ldots, \beta_{s-1},1,\beta_s,\ldots|\gamma)[e^{-i\alpha'\pi\ell_1\cdot k_1}]+\cr
\sum_{r=1}^{|\gamma|} \Big(\prod_{s=1}^r e^{-i\alpha' \pi k_1\cdot k_{\gamma_s}}\Big)
\cA^{(2)}(\alpha|\beta|\ldots,
\gamma_{s-1},1,\gamma_s,\ldots)[e^{-i\alpha'\pi\ell_2\cdot k_1}]\cr
=0\,.
  \end{multline}
At four points we get 
\begin{multline}
\cA^{(2)}(1234)+ e^{i\pi\alpha' k_1\cdot k_2}\! \cA^{(2)}(2134)
+e^{i\pi\alpha'
  k_1\cdot k_{23}}\!\cA^{(2)}(2314)+\cr
 \cA^{(2)}(234|1|.)[e^{-i\pi\alpha'\ell_1\cdot k_1}]
+  \cA^{(2)}(234|.|1)[e^{-i\pi\alpha'\ell_2\cdot k_1}]=0
\label{eq:monod-4pt-2loop}
\end{multline}
where $\cA^{(2)}(1234)$ etc. are planar two-loop amplitude integrand, and
$\cA^{(2)}(234|1|.),\cA^{(2)}(234|.|1)$ are the two non-planar
amplitude integrands with the external state 1 on the $b_I$-cycle with $I=1,2$, as fig.~\ref{fig:two-loop}.
The field theory limit of that relation, at leading order in $\alpha'$, leads to
\begin{multline}
A^{(2)}(1234)+ \cA^{(2)}(2134)
+A^{(2)}(2314)+\cr
A^{(2)}(234|1|.)
+A^{(2)}(234|.|1)=0\,,
\label{eq:monod-4pt-2loop-FT}
\end{multline}
where  $A^{LC}_4(\cdots)$ are the leading colour field theory single trace
amplitudes, and with our
choice of orientation of the cycles $A^{(2)}(234|1|.)
+A^{(2)}(234|.|1)= A_{3;1}(234;1)$ is the double trace field theory amplitude.
We recover the relation obtained by unitarity method
in~\cite{Feng:2011fja}. 
For $\mathcal N=4$ SYM, the graphs are essentially scalar planar and
non-planar double boxes~\cite{Bern:1997nh}, and this relation is
easily verified by inspection, thanks to the antisymmetry of the
three-point vertex. At order $\alpha'$, we conjecture that the field theory
limit yields;
\begin{multline}
k_1\cdot k_2\,A^{(2)}(2134)
+k_1\cdot (k_2+k_3)A^{(2)}(2314)\cr
-A^{(2)}(234|1|.)[\ell_1\cdot k_1]-A^{(2)}(234|.|1)[\ell_2\cdot k_1]=0\,.
\label{eq:monod-4pt-2loop-FT-2}
\end{multline}
These relations are not reducible to KK-like colour relations, like
these of~\cite{Naculich:2011ep}, just like at tree-level where BCJ
kinematic relation go beyond KK ones.  An extension of the one-loop
argument~\cite{Tourkine:2012ip} indicates that the massive string
corrections to the field theory limit of the propagator does not
contribute at the first order in $\alpha'$.  A detailed verification
of this kind of identities will be provided somewhere else, but we
give below a motivation by considering the two-particle discontinuity
in the case of $\mathcal N=4$ SYM. The two-particle $s$-channel cut of
the two-loop amplitude is the sum of two contributions, with one-loop
and tree-level amplitudes, $A(\cdots)$ and
$A^{\rm tree}(\cdots)$~\cite{Bern:1998ug}, respectively:
\begin{multline}
\textrm{disc}_s A^{(2)}(2134)= A(\ell, 21 ,-\tilde\ell) A^{\rm tree}(-\ell ,34,
\tilde \ell)\cr+ A^{\rm tree}(\ell, 21 ,-\tilde\ell) A(-\ell, 34,
\tilde \ell)
\end{multline} where $\ell$ and $\tilde\ell$ are the on-shell cut loop momenta.
The $s$-channel two-particle cut of~\eqref{eq:monod-4pt-2loop-FT-2}
gives a first contribution
\begin{multline}
\hspace{-.8cm}  \left(k_1\cdot \ell_1\, A^{\rm tree}(\ell_1,12,-\tilde \ell_1)+k_1\cdot(\ell_1+k_2)
    A^{\rm tree}(\ell_1,21,-\tilde\ell_1)\right) \cr\times A(-\ell_1 ,34,\tilde\ell_1) =0
\end{multline}
where $\ell_1$ and
$\tilde\ell_1$ are  the cut momenta.
This expression vanishes thanks to the 
monodromy relation between the four-point tree amplitudes in the parenthesis~\cite{Bern:2008qj,BjerrumBohr:2009rd,Stieberger:2009hq}.
The second contribution is 
\begin{multline}
  \Big( A(1,\ell_2,2,-\tilde \ell_2)[k_1\cdot\ell_1]+
    A(\ell_2,12,-\tilde\ell_2)[k_1\cdot(\ell_1+\ell_2)]+\cr
A(\ell_2,21,-\tilde\ell_2)[k_1\!\cdot\!(\ell_1\!+\!\ell_2\!+\!k_2)]
\Big) A^{\rm tree}(-\ell_2 ,34,\tilde\ell_2) =0
\end{multline}
where $\ell_1$ is the one-loop loop momentum and $\ell_2$ and
$\tilde\ell_2$ are  the cut momenta.
This expression vanishes thanks to the four-point one-loop monodromy
relation~\eqref{eq:n-pt-planar} in the parenthesis.
We believe that this approach has the advantage of fixing some
ambiguities in the definition of loop momentum in quantum field
theory.  And the implications of the monodromy relations at higher-loop in maximally
supersymmetric Yang-Mills,  by  applying our construction to the  world-line
formalism of~\cite{Dai:2006vj}, will be studied elsewhere.

Finally, we note that our construction should applies to both the bosonic or
supersymmetric string, as far as the difficulties concerning the
integration of the supermoduli~\cite{Witten:2012bh} can be put aside.

\vspace{-.5cm}
\onecolumngrid
\section*{Acknowledgments}
We would like to thank Lance Dixon for discussions and Tim Adamo, Bo Feng, Michael
B. Green, Ricardo Monteiro, Alexandre Ochirov, Arnab Rudra  for useful
comments on the manuscript. 

The research of PV has received funding the ANR
grant reference QST 12 BS05 003 01, and the CNRS grants PICS number
6430. 
PV is partially supported by   a fellowship funded by the French
Government at Churchill College, Cambridge. The work of PT is supported by STFC grant ST/L000385/1.
The authors would like to thank the Isaac Newton Institute for
Mathematical Sciences, Cambridge, for support and hospitality during
the programme ``Gravity, Twistors and Amplitudes'' where work on this paper was undertaken. This work was supported by EPSRC grant no EP/K032208/1.

\vspace{-.5cm}
\appendix*
\section{Planar and non-planar Green function}
\label{sec:appendixTheta}

The Green function between two external states on the same  boundary of
the annulus $\Ree(\nu_r)=\Ree(\nu_s)$ is given by $\tilde
G(\nu_r,\nu_s)= -\log \vartheta_1(i \Imm(\nu_r-\nu_s)|\tau)/\vartheta_1'(0)$
with $\log q=-2\pi t$
\begin{equation}
  {\vartheta_1( \nu|\tau)\over \vartheta_1'(0)}=   {\sin(\pi \nu)\over \pi}
                                               \prod_{n\geq1} {1-2q^n
                                               \cos(2\pi \nu) +q^{2n}\over                                               (1-q^n)^2}
\end{equation}
and between two external states on the different  boundaries of
the annulus $\Ree(\nu_r)=\Ree(\nu_s)+\frac12$  is given by $\tilde
G(\nu_r,\nu_s)= \log \vartheta_1(\nu_r-\nu_s|\tau)= -\log
\vartheta_2(i\Imm(\nu_r-\nu_s)|\tau)/\theta_1'(0)$
thanks to the relation between the $\vartheta$ functions under the shift $\nu\to \nu+\frac12$ 
\begin{align}
  \label{eq:theta1-half-period}
\vartheta_1(\nu+\frac12|\tau) = \vartheta_2(\nu|\tau),\qquad 
\vartheta_2(\nu+\frac12|\tau)=-\vartheta_1(\nu|\tau)
\end{align}
where
\begin{equation}
  {\vartheta_2( \nu|\tau)\over \vartheta_1'(0)}=
  {\cos(\pi \nu)\over \pi}
                                               \prod_{n\geq1} {1+2q^n
                                               \cos(2\pi \nu) +q^{2n}\over (1-q^n)^2}
\end{equation}
The periodicity around the loop follows from
\begin{align}
  \label{eq:theta1-tau}
  \vartheta_1(\nu+\tau|\tau) =-e^{-i\pi\tau-2i\pi\nu}\,
    \vartheta_1(\nu|\tau); \qquad
\vartheta_2(\nu+\tau|\tau)=e^{-i\pi\tau-2i\pi\nu}\,\vartheta_2(\nu|\tau)\,,
\end{align}
and an appropriate redefinition of the loop momentum. 

The string theory correction $\delta_\pm(x)$ to the field theory
propagator in~\eqref{e:Gdelta} is
 \begin{equation}
  \delta_\pm(x)= -\log\left(1\pm e^{-2i \pi |x|t}\right)\,.  
\end{equation}
$\delta_-(x)$ is the contribution of
  massive string modes propagating between two external states on the
  same boundary  and   $\delta_+(x)$ on different boundaries.

\twocolumngrid   

\bibliography{biblio}

\end{document}